\documentclass[twocolumn,aps,showpacs,floatfix,prc]{revtex4}
\usepackage[dvips]{epsfig}
\begin{document}

\title{ Photonuclear reactions of actinide and pre-actinide nuclei at intermediate energies }

\author{ Tapan Mukhopadhyay$^1$ and D.N. Basu$^2$ }
\affiliation{Variable  Energy  Cyclotron  Centre, 1/AF Bidhan Nagar, Kolkata 700 064, India }

\email[E-mail 1: ]{tkm@veccal.ernet.in}
\email[E-mail 2: ]{dnb@veccal.ernet.in}

\date{\today }

\begin{abstract}

    Photonuclear reaction is described with an approach based on the quasideuteron nuclear photoabsorption model followed by the process of competition between light particle evaporation and fission for the excited nucleus. Thus fission process is considered as a decay mode. The evaporation-fission process of the compound nucleus is simulated in a Monte-Carlo framework. Photofission reaction cross sections are analysed in a systematic manner in the energy range $\sim$~50-70 MeV for the actinides $^{232}$Th, $^{233}$U, $^{235}$U, $^{238}$U and $^{237}$Np and the pre-actinide nuclei $^{208}$Pb and $^{209}$Bi. The study reproduces satisfactorily well the available experimental data of photofission cross sections at energies $\sim$~50-70 MeV and the increasing trend of nuclear fissility with the fissility parameter $Z^2/A$ for the actinides and pre-actinides at intermediate energies [$\sim$~20-140 MeV].
\end{abstract}

\maketitle

\noindent
\section{Introduction}
\label{section1}

    In recent years the study of photofission has attracted considerable interest. When a gamma (photon) above the nuclear binding energy of an element is incident on that element, fission may occur along with the emission of neutrons and the number of neutrons released by each fission event is dependent on the element. With high enough photon energy it is possible to induce fission in most elements. For lighter elements the cross sections are quite small. There is also the difficulty in distinguishing fission from various fragmentation processes, particularly for light elements. These reasons practically limit all investigations to the upper half of the Periodic Table. Recently the experimental techniques are improved by using the solid state track and parallel plate avalanche detectors which make it possible to study fission with very low yields. Moreover, the production of highly neutron-rich radioactive nuclei through photofission will open new perspectives to explore very exotic nuclei far away from the valley of stability, especially in the vicinity of $^{78}$Ni.  

    The use of the energetic electrons is a promising mode to get intense neutron-rich ion beams \cite{Es80}. The reason is that although the fission cross section at the giant dipolar resonance energy is $\approx$~0.16 barn for $^{238}$U (against $\approx$~1.6 barn for 40 MeV neutrons induced fission) but the electrons/$\gamma$-photons conversion efficiency is much more significant than that of the deuterons/neutrons. The energetic beam ($\sim$~50 MeV ) of incident electrons can be slowed down in a tungsten (W) converter or directly in the target (U), generating Bremsstrahlung $\gamma$-rays which may induce fission. In the photofission method for the (neutron-rich) radioactive ion beam production, nuclei are excited by photons covering the peak of the giant dipolar resonance. 

    Isotopes of plutonium and other actinides tend to be long-lived with half-lives of many thousands of years, whereas radioactive fission products tend to be shorter-lived (most with half-lives of 30 years or less). Many of the actinides are very radiotoxic because they also have long biological half-lives and are $\alpha$ emitters as well. In transmutation the intention is to convert the actinides into fission products. The fission products are very radioactive, but the majority of the activity will decay away within a short time. From a waste management viewpoint, transmutation of actinides eliminates a very long-term radioactive hazard and replaces it with a much shorter-term one. Accelerated radioactive decay has been proposed by bombarding spent fuel with electromagnetic (photon) rays. The $\gamma$-rays have also been suggested to be very effective, but $\gamma$-rays are very difficult to produce and may need to be precisely tuned to the target actinide or fission product. 

    The aim of the present work is to obtain photofission cross section. From the photonuclear reactions, it is possible to gain insight into the fission mechanism, especially about how the barrier varies. Information can also be obtained about the excitation of a nucleus by absorption of a high-energy quantum and the subsequent de-excitation process. At intermediate energies [$\sim$~20-140 MeV] such a reaction can be described as a two step process. The incoming photon is assumed to be absorbed by a neutron-proton [n-p] pair inside nucleus, followed by a mechanism of evaporation-fission competition. Photofission cross sections of various nuclei are calculated using a Monte-Carlo method for the evaporation-fission calculation and compared with the available experimental data at energies $\sim$~50-70 MeV. 

\noindent
\section{The photoabsorption mechanism}
\label{section2}

    The dominant mechanism for nuclear photoabsorption at intermediate energies is described by the quasideuteron model \cite{Le51} which is employed to evaluate the total photoabsorption cross section in heavy nuclei. It is based on the assumption that the incident photon is absorbed by a correlated n-p pair inside the nucleus, leaving the remaining nucleons as spectators. Such an assumption is enforced when wavelength of the incident photon is small compared to the nuclear dimensions. The total nuclear photoabsorption cross section $\sigma_a^T$ is then proportional to the available number of n-p pairs inside nucleus and also to the free deuteron photodisintegration cross section $\sigma_d(E_{\gamma})$ and is given by :

\begin{equation}
 \sigma_a^T = \frac{L}{A} NZ \sigma_d(E_{\gamma}) f_P(E_{\gamma})
\label{seqn1}
\end{equation}
\noindent
where $E_{\gamma}$ is the incident photon enegy, $N$, $Z$ and $A$ are the neutron, proton and mass numbers respectively, $L/A$ factor represents the fraction of correlated n-p pairs. Thus, Levinger's constant \cite{Le51} $L$ is a factor which measures the relative probability of two nucleons being near each other in a complex nucleus compared with that in a free deuteron. The function $f_P(E_{\gamma})$ accounts for the reduction of the n-p phase space due to the Pauli exclusion principle. A systematic study of the total nuclear photoabsorption cross section data \cite{Te89} in the intermediate energy range shows that 

\begin{equation}
 f_P(E_{\gamma}) = e^{-D/E_{\gamma}} ~~{\rm where}~~ D = 0.72 A ^{0.81} {\rm MeV}.
\label{seqn2}
\end{equation}
\noindent
For photon energies upto the pion threshold, the Eq. (2) along with the damping parameter $D$ provided above agrees reasonably well with the approach based upon phase space considerations \cite{Ch91} using Fermi gas state densities that conserve linear momentum for the Pauli blocking effects in the quasideuteron regime of hard photon absorption.

    The free deuteron photodisintegration cross section \cite{Wu77} is given by

\begin{equation}
  \sigma_d(E_{\gamma}) = \frac{61.2 ~ [E_{\gamma} - B]^{3/2}}{E_{\gamma}^3} ~ [{\rm mb}]
\label{seqn3}
\end{equation}   
\noindent
where B=2.224 MeV is the binding energy of the deuteron. The quasideuteron model of nuclear photoabsorption is used together with the modern root-mean-square radius data to obtain Levinger's constant 

\begin{equation}
  L = 6.8 - 11.2 A^{-2/3} + 5.7 A^{-4/3}
\label{seqn4}
\end{equation}   
\noindent
of nuclei throughout the Periodic Table and is in good agreement \cite{Ta92} with those obtained from the experimentally  measured $\sigma_a^T$ values. At the quasideuteron energy range, as a consequence of the primary photointeraction, $\gamma$+(n+p)$\rightarrow$n*+p*, in most of the cases excited compound nuclei are formed with the same composition as target nucleus where both neutron and proton are retained inside the nucleus and the probabilities that either neutron escapes or proton escapes or both neutron and proton escape from the nucleus are extremely low.

\noindent
\section{The nuclear excitation and fission probability}
\label{section3}

    After absorption of a photon ($\gamma$) with incident energy $E_\gamma$ (as measured in the laboratory frame), the nucleus with rest mass $m_0$ recoils with a velocity $v_r$ given by 

\begin{equation}
 v_r= \frac{E_{\gamma}c}{[E_{\gamma}+m_0 c^2]}
\label{seqn5}
\end{equation}   
\noindent
which is also the velocity of the centre of mass $v_{cm}$ of the $\gamma$-nucleus system where $c$ is the speed of light in vacuum. In the centre of mass frame, momenta of the $\gamma$ and the nucleus are the same and equal to $p_{cm}$:

\begin{equation}
 p_{cm}= \frac{E_{\gamma}m_0c}{\sqrt{[m_0c^2(2E_{\gamma}+m_0 c^2)]}}
\label{seqn6}
\end{equation}   
\noindent
and the kinetic energy $E_r$ of the recoiling nucleus in the laboratory frame is

\begin{eqnarray}
 E_r=&& \frac{m_0^{\prime}c^2}{\sqrt{1-v_r^2/c^2}}-m_0^{\prime}c^2 \nonumber \\
      =&&E_{\gamma}+m_0c^2-[m_0c^2(2E_{\gamma}+m_0c^2)]^{1/2}
\label{seqn7}
\end{eqnarray}   
\noindent
where  $m_0$ and $m_0^{\prime}$ are the rest masses of the nucleus before and after the photon absorption respectively. 

    The recoiling nucleus can be viewed as a compound nucleus having the same composition as the target nucleus but 
with the excitation energy 

\begin{eqnarray}
 E^*=&&m_0^{\prime}c^2-m_0 c^2 = m_0 c^2 [(1 + 2E_{\gamma}/m_0 c^2 )^{1/2} - 1] \nonumber \\
      =&&[m_0c^2(2E_{\gamma}+m_0c^2)]^{1/2}-m_0c^2
\label{seqn8}
\end{eqnarray}   
\noindent
and in this case $E^*$ is also equal to the kinetic energy in the centre of mass frame (which is sum of the kinetic energies of the $\gamma$ and the nucleus moving in the centre of mass frame). The kinetic energy $E_r$ of the recoiling nucleus in the laboratory frame can now be rewritten as the obvious result

\begin{equation}
 E_r= E_{\gamma} - E^*.
\label{seqn9}
\end{equation}   
\noindent
This excited compound nucleus then undergoes successive evaporation of neutrons and other light particles or fission. Thus, the fission is considered as a decay mode. The photofission cross section $\sigma_f$ is a product of the nuclear photoabsorption cross section $\sigma_a^T$ and the total fission probability (fissility) $f$ and is, therefore, given by 

\begin{equation}
 \sigma_f=\sigma_a^T f.  
\label{seqn10}
\end{equation}                                                                                                                                           
\noindent    

    The statistical approach for nucleon and light-particle evaporation and nuclear fission is an appropriate scheme for calculation of the relative probabilities of different decay modes of the compound nucleus. Such statistical decay of the compound nucleus is the slow stage of the photonuclear reaction. According to the standard Weisskopf evaporation
scheme \cite{We37}, the partial width $\Gamma_j$ for the evaporation of a particle $j$ = n, p, $^2$H, $^3$H, $^3$He or $^4$He is given by

\begin{equation}
 \Gamma_j = \frac{(2s_j+1)\mu_j}{\pi^2\hbar^2\rho_{CN}(E^*)} \int_{V_j}^{E^*-B_j} \sigma^j_{inv}(E)\rho_j(E^*-B_j-E) EdE
\label{seqn11}
\end{equation}
\noindent
where $s_j$, $\mu_j$, $V_j$ and $B_j$ are the spin, reduced mass, Coulomb barrier and the binding energy of the particle $j$, respectively. $\sigma^j_{inv}(E)$ is the cross section for the inverse reaction which means the capture reaction cross section of the particle $j$ to create the compound nucleus. $\rho_{CN}$ and $\rho_j$ are the nuclear level densities for the initial and final (after the emission of the particle $j$) nuclei, respectively.

    The Bohr-Wheeler statistical approach \cite{Bo39} is used to calculate the fission width of the excited compound nucleus. This width is proportional to the nuclear level density $\rho_f$ at the fission saddle point:

\begin{equation}
 \Gamma_f = \frac{1}{2\pi\rho_{CN}(E^*)} \int_{0}^{E^*-B_f} \rho_f(E^*-B_f-E) dE
\label{seqn12}
\end{equation}
\noindent
where $B_f$ is the fission barrier height. The diffuse surface nucleus Sierk's \cite{Si79} fission barriers ($B_f$) are used for these calculations. The decay of the excited compound nucleus \cite{Ha52} is simulated using the Monte-Carlo method \cite{Pa84}. The competition between the various decay channels at each step of the evaporation chain is determined by the relation between the partial widths for particle evaporation and fission, Eqs. (11) and (12), respectively. Finally, in order to calculate the fission probability, the total number of fission events in a computer run is counted and divided by the total number of photoabsorption events. Evaporation from excited fission fragments is also taken into account. 

\noindent
\section{Competition between light particle evaporation and fission}
\label{section4}

    The quantitative description of the process is based on the liquid drop model (LDM) for nuclear fission by Bohr and Wheeler \cite{Bo39} and the statistical model of nuclear evaporation developed by Weisskopf \cite{We37}. Accordingly, the probability of fission relative to neutron emission is calculated using Vandenbosch-Huizenga's equation \cite{Hu73}  given by

\begin{equation}
 \frac{\Gamma_f}{\Gamma_n} = \frac{K_0 a_n[2(a_f E_f^*)^\frac{1}{2}-1]}{4A^\frac{2}{3}a_f E_n^*} \exp{{\big [}2 [(a_f E_f^*)^\frac{1}{2} - (a_n E_n^*)^\frac{1}{2}] {\big ]}}
\label{seqn13}
\end{equation}
\noindent
where $E_n^*=E^*-B_n$ and $E_f^*=E^*-B_f$ are the nuclear excitation energies after the emission of a neutron and after fission, resepctively, where $B_n$ is the binding energy of the emitted neutron. $\Gamma_n$ and $\Gamma_f$ are the partial widths for the decay of the excited compound nucleus via neutron emission and fission, resepctively, and the parameters $a_n$ and $a_f$ are the level density parameters for the neutron emission and the fission, respectively and $K_0=\hbar^2/2mr^2_0$ where $m$ and $r_0$ are the neutron mass and radius parameter respectively. The emission probability of particle $k$ relative to neutron emission is calculated according to the Weisskopf's statistical model \cite{We37}

\begin{equation}
 \frac{\Gamma_k}{\Gamma_n} = {\Big (}\frac{\gamma_k}{\gamma_n}{\Big )} {\Big (}\frac{E_k^*}{E_n^*}{\Big )}  {\Big (}\frac{a_k}{a_n}{\Big )} \exp{{\big [}2 [(a_k E_k^*)^\frac{1}{2} - (a_n E_n^*)^\frac{1}{2}] {\big ]}}
\label{seqn14}
\end{equation}
\noindent
where $a_k$ is the level density parameter for the emission of the particle $k$, $\gamma_k/\gamma_n =1$ for $k=p$, 2 for $k=^4$He, 1 for $k=^2$H, 3 for $k=^3$H and 2 for $k=~^3$He. $E_k = E^* - (B_k + V_k)$ is the nuclear excitation energy after the emission of particle $k$ \cite{Le50}. $B_k$ are the binding energies of the emitted particles and $V_k$ are the Coulomb potentials. The evaporation-fission competition of the compound nucleus is then described in a Monte-Carlo framework. Any particular reaction channel $r$ is then defined as the formation of the compound nucleus via photoabsorption and its decay via particle emission or fission. Thus, fission is considered as a decay mode. The photonuclear reaction cross sections $\sigma_r$ are calculated using the equation $\sigma_r = \sigma_a^T n_r/N$ where $n_r$ is the number of events in a particular reaction channel $r$ and $N$ is total number of events that is the number of the incident photons. 

\begin{table}[htbp]
\vspace{-0.07cm}
\caption{\label{tab:table1} Comparison between the measured and the calculated photofission reaction cross sections.}
\begin{tabular}{|c|c|c|c|c|}
\hline
Target&$E_{\gamma}$&Expt.        &Calc.&Calc.          \\ 
nuclei&                    &$\sigma_f$&$\sigma_a^T$  &$\sigma_f$\\
           & MeV &    mb [Ref.]     & mb & mb            \\  
\hline
$^{237}$Np&52&20$\pm$2 \cite{Iv89}&17.6&17.61$\pm$ 0.09\\ 
&69&19$\pm$4 \cite{Iv89}&15.6&15.59 $\pm$0.05\\ 
\hline
$^{235}$U&52&16$\pm$4 \cite{Le87}&17.6&17.58$\pm$ 0.09\\ 
&69&15$\pm$3 \cite{Le87}&15.5&15.54 $\pm$0.08\\ 
\hline
$^{238}$U&52 &14$\pm$2 \cite{Iv89}&17.5&17.46$\pm$ 0.09\\ 
&&16$\pm$2 \cite{Le87}&& \\
&&32$\pm$2 \cite{Mo69}&& \\
&&25$\pm$3 \cite{Be88}&& \\
&69 &15$\pm$1 \cite{Le87}&15.5&15.52$\pm$ 0.08\\ 
&&23$\pm$1 \cite{Le87}&& \\
&&13$\pm$2 \cite{Iv89}&& \\
\hline
$^{232}$Th&52 &8.6$\pm$0.6 \cite{Le87}&17.5&12.41 $\pm$0.07\\ 
&69& 9$\pm$1 \cite{Le87}&15.4&13.82 $\pm$0.07\\ 
\hline
$^{209}$Bi&52 &(16$\pm$1)$\times$10$^{-3}$ \cite{Mo69}&17.4&0.15 $\pm$0.01\\ 
&&(70$\pm$12)$\times$10$^{-3}$ \cite{Ar86}&& \\
&&(24$\pm$3)$\times$10$^{-3}$ \cite{Le80}&& \\
&69 &(8.0$\pm$0.6)$\times$10$^{-2}$ \cite{Mo69}&15.0&0.26 $\pm$0.01\\ 
&&(18$\pm$3)$\times$10$^{-2}$ \cite{Ar86}&& \\
                 &120&0.20$\pm$0.06 \cite{Te98}&9.3&0.87 $\pm$0.01 \\ 
                 &145&0.31$\pm$0.08 \cite{Te98}&7.6&1.21 $\pm$ 0.02\\ 
\hline
$^{208}$Pb&52&(1.9$\pm$0.3)$\times$10$^{-3}$ \cite{Mo69}&17.3&0.039 $\pm$0.004\\ 
&&3$\times$10$^{-3}$ \cite{Ar90}&& \\
&69 &(12$\pm$2)$\times$10$^{-3}$ \cite{Mo69}&14.9&0.097$\pm$0.006 \\ 
&&(18$\pm$3)$\times$10$^{-3}$ \cite{Ar90}&& \\
\hline
$^{197}$Au&120&(59$\pm$38)$\times$10$^{-3}$ \cite{Te98}&9.0&0.073 $\pm$0.004\\ 
                 &145&(11$\pm$4)$\times$10$^{-2}$ \cite{Te98}&7.3&0.117$\pm$0.005\\ 
\hline
$^{181}$Ta&120&(13$\pm$4)$\times$10$^{-3}$ \cite{Te98}&8.5&0.0023$\pm$0.0007 \\ 
                 &145&(9.7$\pm$3.0)$\times$10$^{-3}$ \cite{Te98}&6.9&0.0035$\pm$0.0008\\ 
\hline
$^{174}$Yb&52 &(3.2$\pm$0.5)$\times$10$^{-5}$ \cite{Mo69}&16.7&$<$~4.2$\times$10$^{-4}$ \\ 
&69 &(6$\pm$1)$\times$10$^{-5}$ \cite{Mo69}&13.9&$<$~3.5$\times$10$^{-4}$ \\ 
\hline
$^{154}$Sm&69 &(1.8$\pm$4)$\times$10$^{-7}$ \cite{Mo69}&13.0&$<$~3.3$\times$10$^{-4}$ \\ 
\hline
$^{51}$V&120&(76$\pm$25)$\times$10$^{-3}$ \cite{Te98}&3.0&0.0019$\pm$0.0004 \\ 
                 &145&(78$\pm$29)$\times$10$^{-3}$ \cite{Te98}&2.3&0.0033 $\pm$0.0004\\ 
\hline
\end{tabular} 
\vspace{-.07cm}
\end{table}
         
\noindent
\section{Comparison of the photofission cross section estimates with measured data}
\label{section5}

    Each calculation is performed with 40000 events using a Monte-Carlo technique for the evaporation-fission calculation. This provides a reasonably good computational statistics. The photofission cross sections are calculated at different energies for various elements. Results of these calculations corresponding to the available experimental data at 52 MeV, 69 MeV, 120 MeV and 145 MeV are listed in Table-I. The statistical error in the theoretical estimates for the photofission cross sections are calculated using the equation $\sigma_f \pm \Delta\sigma_f= \sigma_a^T [n_f \pm \sqrt{n_f}]/N$ which implies that  $\Delta\sigma_f=\sqrt{\sigma_a^T\sigma_f/N}$. For the cases, where not a single fission event occured in N events, the upper limits of the cross sections are calculated using the equation $\sigma_f = \sigma_a^T/N$ where $N [=40000]$ is the number of incident photons.

\begin{figure}[htbp]
\vspace{0.4cm}
\eject\centerline{\epsfig{file=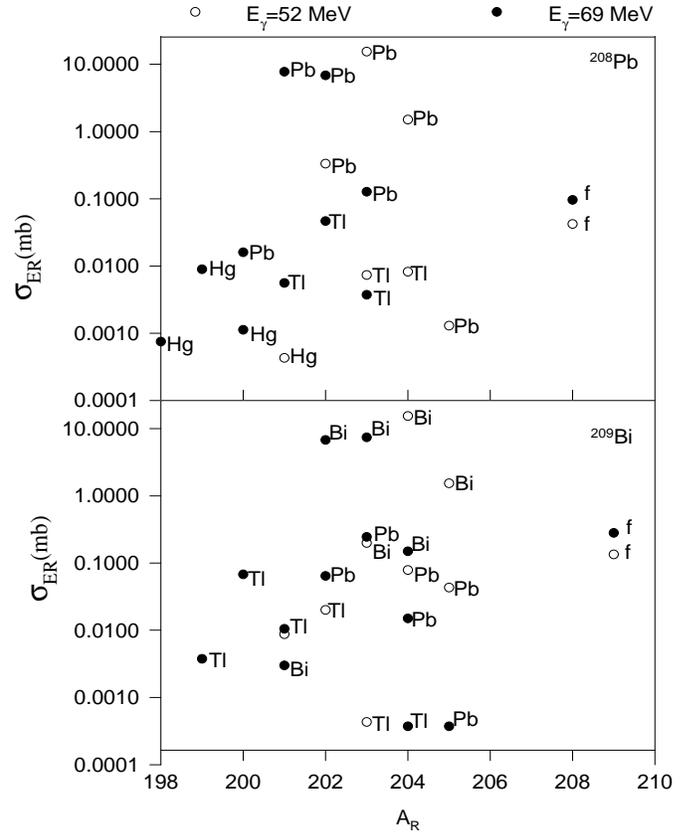,height=11cm,width=8.8cm}}
\caption
{The plots of cross sections $\sigma_{ER}$ as a function of mass number $A_R$ of the evaporation residues for the pre-actinides $^{208}$Pb and $^{209}$Bi at $E_{\gamma}=$ 52 MeV, 69 MeV. The fission cross sections of the fissioning nuclei are also shown (marked $f$).}
\label{fig1}
\vspace{0.4cm}
\end{figure}
\noindent

    The results of the present calculations should be compared with measured photofission cross sections with certain degree of caution. In this regard, it is pertinent to mention here that the experimental data listed in Table-I for $E_{\gamma}=$ 69 MeV, 120 MeV and 145 MeV correspond to the effective mean energies of 69 MeV, 120 MeV and 145 MeV respectively, for the incident quasi-monochromatic beams of photons. However, those for $E_{\gamma}=$ 52 MeV actually correspond to the incident beam energy of 52 MeV monochromatic photons. The same holds for the cross sections for the evaporation residues calculated for these photon induced reactions. But the theoretical calculations are performed exactly at photon energies of 52 MeV, 69 MeV, 120 MeV and 145 MeV. Nevertheless, present calculations provide good estimates of the photofission cross sections for the actinides. In the pre-actinide to medium mass region the agreement with the measured photofission cross sections is reasonable. 
                
\begin{table*}[htbp]
\vspace{-0.07cm}
\caption{\label{tab:table1} Variation of the calculated photofission reaction cross sections of actinides with the incident photon energy ($E_{\gamma}$) and the fissility parameter $Z^2/A$.}
\begin{tabular}{cccccccccc}
\hline
\hline
Target&$Z^2/A$&Calculated&$E_{\gamma}$[MeV]&$E_{\gamma}$ [MeV]&$E_{\gamma}$ [MeV]&$E_{\gamma}$ [MeV]&$E_{\gamma}$ [MeV]&$E_{\gamma}$ [MeV]&$E_{\gamma}$ [MeV]      \\ 
nuclei&   &Quantity&20 &40 &60 &80 &100 &120 &140  \\
\hline 
\hline
$^{239}$Pu&36.97&$\sigma_a^T$ (mb)&10.19 &10.04 &16.78 &14.24 &11.97 &10.13 &8.68 \\
                 &&$\sigma_f$ (mb)   &10.19 &10.04 & 17.78&14.24 &11.97 & 10.13& 8.68\\
                 &&$f$~($\%$)          & 99.98&100.0 &100.0 &100.0 &100.0 &100.0  &99.99\\
\hline
$^{237}$Np&36.49&$\sigma_a^T$ (mb)&10.31 &18.06 &16.74 &14.18 &11.90 &10.07 &8.62 \\
                 &&$\sigma_f$ (mb)   &10.24 &18.05 &16.73 &14.18 &11.90 &10.07 & 8.62\\
                 &&$f$~($\%$)          & 99.38&99.99 &99.99 &99.99 &100.0 & 99.99 &99.99 \\
\hline
 $^{233}$U&36.33&$\sigma_a^T$ (mb)&10.58 &10.15 &16.71 & 14.11&11.82 &9.99 &8.54 \\
                 &&$\sigma_f$ (mb)   &10.40&10.14 &16.70 &14.11 &11.81 &9.98 &8.53 \\
                 &&$f$~($\%$)          &98.34 &99.95 &99.97 &99.99 &99.97 &99.96  &99.93 \\
\hline
 $^{235}$U&36.02&$\sigma_a^T$ (mb)&10.42 &18.07 &16.69 &14.12 &11.84 &10.01 &8.57 \\
                 &&$\sigma_f$ (mb)   &9.98 &18.05 &16.68 &14.11 &11.84 &10.00 &8.56 \\
                 &&$f$~($\%$)          &95.83 &99.91& 99.97&99.96 & 99.99&99.97  &99.95 \\
\hline
 $^{238}$U&35.56&$\sigma_a^T$ (mb)&10.19 &17.94 &16.66 &14.13 &11.87 & 10.04& 8.60\\
                 &&$\sigma_f$ (mb)   &8.94&17.76 &16.64 &14.12 &11.86 &10.04 &8.59 \\
                 &&$f$~($\%$)          &87.71 &98.97&99.87  &99.96 &99.95 &99.96  &99.97 \\
\hline
$^{232}$Th&34.91&$\sigma_a^T$ (mb)&10.57 &18.05 &16.59 &13.99 &11.72 &9.89 &8.46 \\
                 &&$\sigma_f$ (mb)   &2.20 &10.18 &13.86 &13.27 &11.45 &9.70 &8.27 \\
                 &&$f$~($\%$)          &20.87 &56.38 &83.53 &94.79 &97.76 &98.06  &97.83 \\ \hline
\hline
\end{tabular} 
\vspace{-0.07cm}
\end{table*}

\noindent
\section{Results and discussion}
\label{section6}

    The photonuclear reaction cross section calulations at intermediate energies within a Monte-Carlo framework for simulation of the evaporation-fission competition are performed assuming 40000 incident photons for each calculation which provide a reasonably good computational statistics. The cross sections of fission and evaporation residues ($\sigma_{ER}$) as a function of mass number $A_R$ of the evaporation residues for the pre-actinides $^{208}$Pb and $^{209}$Bi at the incident photon energies of 52 MeV and 69 MeV for the present calculations are shown in Fig.1. This is to show the relative probabilities of other evaporation processes compared to fission. The results of the fission cross sections and the fissility at $E_{\gamma}=$ 20, 40, 60, 80, 100, 120 and 140 MeV using 40000 events for the Monte-Carlo calculations are tabulated in Table-II. The increasing trend of the nuclear fissility with the fissility parameter $Z^2/A$ for the actinides at intermediate energies [$\sim$~20-140 MeV] are observed.

    Present calculations provide excellent estimates of the photofission cross sections for the actinides except for $^{232}$Th where it somewhat overestimates the fission cross sections. However, the general increasing trend of nuclear fissility with fissility parameter is retained. For the pre-actinides, $^{208}$Pb and $^{209}$Bi, the photonuclear reaction cross sections show peaks for evaporation residues around $A_R=$ 203 (Pb), 201 (Pb) and 204 (Bi), 203 (Bi) respectively, for the incident photon energies $E_{\gamma}=$ 52 MeV, 69 MeV. The total number of events (that is the number of incident photons) for each run limits the calculations of too low fission cross-sections. The fissility for thorium \cite{Bi93} and several uranium isotopes \cite{Ce00} was found to be lower than that for neptunium, showing that nuclear fissility does not saturate for those nuclei, remaining at a value below hundred percent even at high incident photon energies. The present theoretical study corroborates this behaviour. We have included $^{239}$Pu in our study, though it is not a naturally occuring element but is readily produced as spent fuel.

\noindent
\section{Summary and conclusion}
\label{section7}

    In summary, the cross sections for the fission and the evaporation residues are calculated for photon induced reactions at intermediate energies. Monte-Carlo calculations for the evaporation-fission competition are performed assuming 40000 incident photons for each calculation. These many number of events for each calculation provide a reasonably good statistics for computationally stable results. Present calculations provide excellent estimates of the photofission cross sections for the actinides. In the pre-actinide to medium mass region the agreement with measured photofission cross sections is reasonable. The process of photofission of heavy nuclei is considered in terms of production of fission fragments and is projected as a viable method for the production of neutron rich nuclei for radioactive ion beam (RIB). 

\noindent

\end{document}